\documentclass[preprint2]{aastex}

\shortauthors{Wright \etal}

\shorttitle{Cool Brown Dwarf}

\newcommand{\etal}         {{\it et al.}}
\newcommand{\vs}           {{\it vs.}}
\newcommand{\asec}	{\mbox{$^{\prime\prime}$}}
\newcommand{\amin}	{\mbox{$^\prime$}}
\newcommand{\um}        {\mbox{$\mu$m}}

\newcommand{\be}           {\begin{equation}}
\newcommand{\ee}           {\end{equation}}
\newcommand{\bea}          {\begin{eqnarray}}
\newcommand{\eea}          {\end{eqnarray}}

\begin{document}

\title{NEOWISE-R Observation of the Coolest Known Brown Dwarf}

\author{
Edward L.\ Wright\altaffilmark{1},
Amy Mainzer\altaffilmark{2},
J.\ Davy Kirkpatrick\altaffilmark{3},
Frank Masci\altaffilmark{3},
Michael C.\ Cushing\altaffilmark{4},
James Bauer\altaffilmark{2},
Sergio Fajardo-Acosta\altaffilmark{3},
Christopher R.\ Gelino\altaffilmark{3},
Charles A.\ Beichman\altaffilmark{3,5},
M.\ F.\ Skrutskie\altaffilmark{6},
T.\ Grav\altaffilmark{7},
Peter R.\ M.\ Eisenhardt\altaffilmark{2},
Roc Cutri\altaffilmark{3}
}

\altaffiltext{1}{UCLA Astronomy, PO Box 951547, Los Angeles CA 90095-1547}
\altaffiltext{2}{Jet Propulsion Laboratory, 4800 Oak Grove Drive, Pasadena, CA 91109}
\altaffiltext{3}{Infrared Processing and Analysis Center, California Institute of Technology, Pasadena CA 91125}
\altaffiltext{4}{Department of Physics and Astronomy, University of Toledo, 2801 W. Bancroft St., Toledo\ OH\ 43606-3328}
\altaffiltext{5}{NASA Exoplanet Science Institute, MS 100-22, California Institute of Technology, Pasadena CA 91125}
\altaffiltext{6}{Department of Astronomy, University of Virginia, Charlottesville, VA, 22904}
\altaffiltext{7}{Planetary Science Institute, Tucson, AZ 85719}
\email{wright@astro.ucla.edu}

\begin{abstract}
The Wide-field Infrared Survey Explorer (WISE) spacecraft has 
been reactivated as NEOWISE-R to characterize and search for Near Earth Objects. 
The brown dwarf WISE J085510.83-071442.5 has now been reobserved by NEOWISE-R,
and we confirm the results of Luhman (2014b),
who found  a very low effective temperature ($\approx 250$ K),
a very high proper motion ($8.1 \pm 0.1\;\asec$/yr) , and a large parallax ($454 \pm 45$ mas).
The large proper motion has separated the brown dwarf from the
background sources that influenced the 2010 WISE data, allowing a
measurement of a very red WISE color of W1-W2 $> 3.9$ mag.  
A re-analysis
of the 2010 WISE astrometry using only the W2 band, combined with the
new NEOWISE-R 2014 position, gives an improved parallax of 
$448\pm 33$ mas and proper motion of $8.08\pm 0.05\;$\asec/yr. These are
all consistent with Luhman (2014b).
\end{abstract}


\maketitle

\section{Introduction}

The Wide-field Infrared Survey Explorer (WISE) 
\citep{wright/etal:2010}
observed the
entire sky in four infrared bands at 3.4, 4.6, 12 and 22 \um\ 
in early 2010, then continued to observe in the 3.4 and 4.6 \um\ 
bands until Feb 2011.  In Feb 2011 the spacecraft was placed 
into hibernation.  In late 2012 a brief test
showed that the spacecraft was still functional, and in 2013
the planetary science division of NASA funded a reactivation
of the spacecraft to characterize and search for Near Earth Objects,
with Amy Mainzer as Principal Investigator of this new
NEOWISE-R (NEO WISE Reactivation) mission.  
After passively cooling back to 73 K, 
surveying the sky began again in December 2013
\citep{mainzer/etal:2014}.

Both \citet{luhman:2014a} and \citet{kirkpatrick/etal:2014}
noted that the source \\
WISE J085510.83-071442.5 \\
(also WISEA J085510.74-071442.5,  \\
hereafter W0855) had a large
motion between the two epochs of WISE data in May and November
2010.  But since the source could not be seen in the 2MASS
survey \citep{skrutskie/etal:2006}, this motion was unconfirmed.
\citet{luhman:2014b} also failed to see W0855 in the J band images
from the VISTA survey \citep{mcmahon/etal:2013}, 
but was able to obtain data using
the short wavelength bands of the IRAC camera \citep{fazio/etal:2004}
on the Spitzer Space Telescope \citep{werner/etal:2004}
that confirmed the large motion and
showed a large parallax as well.  In addition, these data showed that
W0855 was an extremely red source, and that a clump of background
sources had influenced the WISE photometry and astrometry taken in 2010.
The low luminosity and red color of W0855 require effective temperatures near
250 K and masses well below the deuterium
burning limit: 3 to 10 Jupiter masses for ages of 1 to 10 Gyr 
\citep{luhman:2014b}.   Further observations of W0855 and further searches
for similar objects will aid our understanding of the relation between the
population of more massive brown dwarfs and the population of exoplanets
both bound and free-floating \citep{sumi/etal:2011}.

In May 2014 NEOWISE-R scanned over W0855 again, giving new unconfused
position and photometry data for both W0855 and the clump of
background sources.  In this paper we use these data to derive an improved
proper motion and parallax fit for W0855, and to derive uncontaminated
colors in the WISE bands.  Figure \ref{fig:W12} shows
postage stamps of the WISE W1 and W2 coadds for the three epochs.
In addition we report an unsuccessful attempt to detect W0855 in the H band.

\section{Observations}

\begin{figure}[tbh]
\plotone{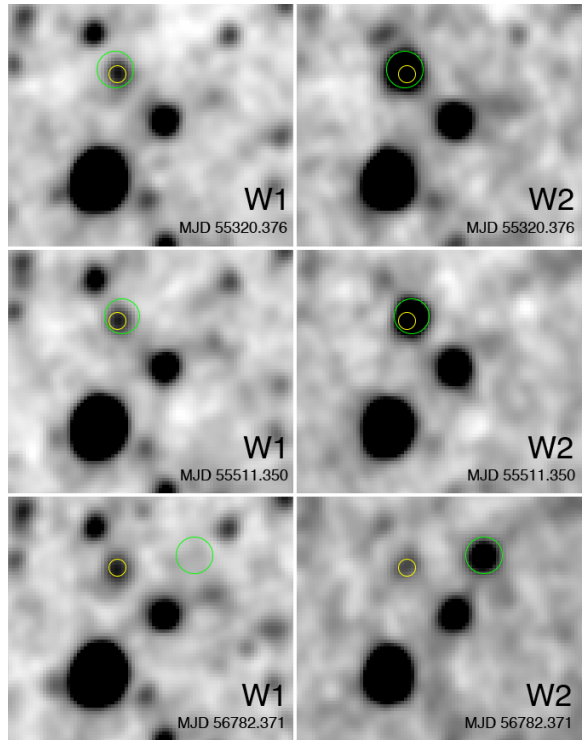}
\caption{W1 and W2 images of the W0855 field for the three WISE epochs.
The small yellow circle marks the background source clump.
The larger green circle marks the WISE position for each
epoch.  Each panel is 100\asec\ wide, with North at the top and
East on the left.
\label{fig:W12}}
\end{figure}

\begin{figure}[tbh]
\plotone{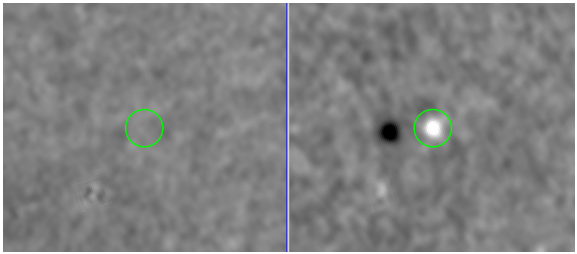}
\caption{May 2014 minus May 2010 images.
Left panel: W1, right panel: W2.  Width of each
image is 3.5\amin.  The circle marks the May
2014 position of W0855.
\label{fig:w1w2diff}}
\end{figure}

\begin{table*}[tb]
\caption{New Astrometric Data [J2000] \label{tbl:astrom}}
\begin{tabular}{rrrrrl}
MJD & $\alpha$ $[{}^\circ]$ & $\sigma_\alpha$ [mas] & $\delta$ $[{}^\circ]$ & $\sigma_\delta$ [mas] & Notes \\
\hline
55320.376 & 133.7951649 & 101 & -7.2451021 & 111 & WISE W2 only \\
55511.350 & 133.7944072  & 109 & -7.2450719 & 120 & WISE W2 only \\
56782.371 & 133.7862181  & 158 & -7.2442562 & 175 & NEOWISE-R \\
\hline
56782.371 & 133.7949499  & 207 & -7.2455978 & 231 & Bkgnd, W1 \& W2 \\
56782.371 & 133.7949398  & 486 & -7.2452772  & 556 & Bkgnd, W2 only \\
\hline
\end{tabular}
\end{table*}

The 2014 NEOWISE-R data are reported in Table \ref{tbl:astrom}
and Table \ref{tbl:phot}.  
These values were all obtained using the AllWISE profile fit
photometry and astrometry software applied to the sets of frames
containing W0855 in each epoch.
The combined flux from W0855
and the background source clump gives a W2 magnitude of
$13.757 \pm 0.050$, which is reasonably close to the 
$13.633 \pm 0.038$ 
observed in May 2010.  
All magnitudes and colors in this paper are reported on the Vega
system.
The W1-W2 color observed in 2010 was
clearly contaminated by the background source clump.
The new color W1-W2 = $3.803 \pm 0.329$ mag is extremely red,
placing W0855 squarely in the region of 
Y dwarfs \citep{kirkpatrick/etal:2012}.  The signal
to noise ratio in W1 is quite low, so the limits on the color are
asymmetric: at $2\sigma$, the range of colors consistent with
the data is  3.3 to 4.8 mag.  Examination of Figure \ref{fig:W12}
shows that the NEOWISE-R W1 flux may still be confused by
a much fainter background source, and that the density of faint background
objects is such that the W1 flux measurement of W0855 is confusion 
limited.

In order to reduce the effects of confusion, difference images
of the May 2014 coadd minus the May 2010 coadd were
constructed.  These are shown in Figure \ref{fig:w1w2diff}.
Analysis of these images shows that the W1 flux of W0855
is $7.5 \pm 6.5\;\mu$Jy.  This is about 2.3$\sigma$ less than
the value found by the standard profile fit photometry,
indicating that the May 2014 epoch probably still suffers from
confusing background sources. 
The central value of the W1-W2 color is 5.0 mag, with a 2$\sigma$
lower limit on the color of 3.9 mag.  W0855 is an extremely red
source.

W0855 was observed on 9 Feb 2014 in the H (1.6 $\mu$m) band with
NIRC2 behind the Keck II LGS-AO system 
\citep{wizinowich/etal:2006,vandam/etal:2006}, but the 40\asec\ field-of-view
was improperly
centered due to the large proper motion of W0855.
W0855 was reobserved on 18 May 2014 but 
no detection was made in 30 minutes of
integration, and a $3\sigma$ upper limit
is reported in Table \ref{tbl:phot}.   This upper limit shows
that W0855 is redder than W1828+2650, one of the reddest known brown
dwarfs,  in the H-W2 color
\citep{kirkpatrick/etal:2012}.  The upper limit $J > 23$
reported by \cite{luhman:2014b} shows that W0855 is
also nearly as red or redder than W1828+2650 in the J-W2 color.

\begin{table}[tbh]
\caption{New Photometric Observations.\label{tbl:phot}}
\begin{tabular}{lll}
\tableline\tableline
Filter & Magnitude & Notes\\
\tableline
W1 & $16.117\pm0.073$ & Background Clump\\
W2 & $15.441\pm0.152$ & Background Clump\\
\tableline
H & $>22.7$ & W0855 \\
W1 & $17.819\pm0.327$ & W0855 \\
W2 & $14.016\pm0.048$ & W0855 \\
\tableline
\end{tabular}
\end{table}

\section{Astrometric Fit\label{s:astfit}}

\begin{figure}[tbh]
\plotone{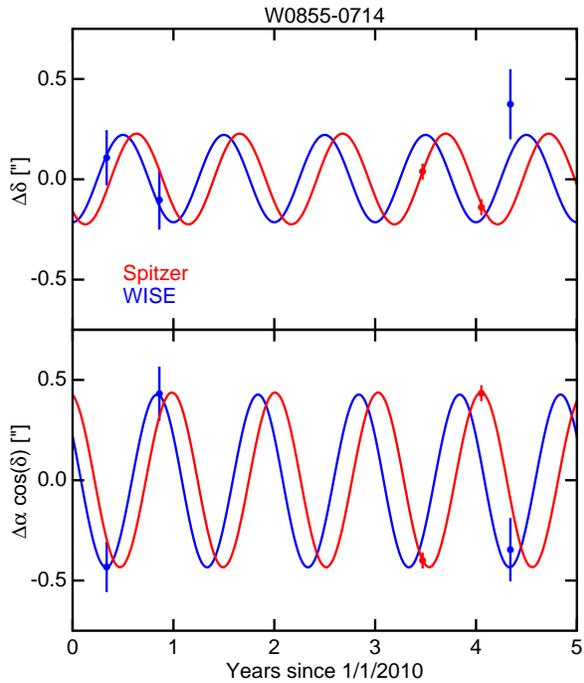}
\caption{Astrometric data and fit for W0855.
The constant and proper motion terms
have been taken out in order to clearly show
the parallactic motion.  The parallax curve for
Spitzer has a different phase due to its 
Earth-trailing orbit.
\label{fig:astrom}}
\end{figure}

Also reported in Table \ref{tbl:astrom} are new analyses of the
2010 WISE data.  For these positions, only the W2 data were
used.  Since the detection of W0855 is dominated by W2 and the
detection of the background source clump is dominated by W1,
the positions will be less effected by confusion.

Even with the positions derived from W2 data alone, we
expect to see an astrometric shift due to the influence of
the background source clump.  \citet{luhman:2014b} went
to considerable lengths to correct for this influence.
In order to check Luhman's analysis using an independent
method, we use a fairly simple analysis where which assumes
that the true position is related to the observed position
by
\be
(\alpha,\delta)_\mathrm{true}
= f\times (\alpha,\delta)_\mathrm{obs}
+ (1-f) \times (\alpha,\delta)_\mathrm{bkgnd}
\ee
We model the true position by the usual five
parameters: $\Delta\alpha_\circ,\;\Delta\delta_\circ,\;
 \mu_\alpha = \cos\delta\;  d\alpha/dt,\; \mu_\delta  = d\delta/dt$ and parallax $\varpi$.
This introduces a sixth parameter $f$ into the analysis
in  addition to the usual five parameters. 
Only the first two observations by
WISE are affected by this new parameter. 

We end up with the standard equations \citep{marsh/etal:2013}
for $i = 3$ to 5:
\bea
x_i & = & \cos\delta_1 (\alpha_i-\alpha_\mathrm{bkgnd}) \nonumber \\
& = & \Delta\alpha + \mu_\alpha (t_i-t_1) + \varpi \vec{R}_i \cdot \hat{W}  \nonumber \\
y_i  & = & \delta_i-\delta_\mathrm{bkgnd}  \nonumber \\
& = & \Delta\delta +\mu_\delta (t_i-t_1) - \varpi \vec{R}_i \cdot \hat{N}
\eea
and modified equations for $i = 1$ and 2:
\bea
x_i & = & \cos\delta_1 (\alpha_i-\alpha_\mathrm{bkgnd}) \nonumber \\
& = & \Delta\alpha + \mu_\alpha (t_i-t_1) + \varpi \vec{R}_i \cdot \hat{W} - (f-1)x_i \nonumber \\
y_i & = & \delta_i-\delta_\mathrm{bkgnd}   \nonumber \\
& = & \Delta\delta +\mu_\delta (t_i-t_1) - \varpi \vec{R}_i \cdot \hat{N} - (f-1)y_i\nonumber \\
& & 
\label{eqn:modeq}
\eea
\noindent where $t_i$ is the observation time [yr] of the $i^{th}$ astrometric
measurement, and $R_i$ is the vector
position of the observer relative to the Sun in celestial
coordinates and astronomical units. $\hat{N}$ and $\hat{W}$
are unit vectors pointing North and West from the position of the source.
Epochs 3 \& 4 are the Spitzer positions reported by \citet{luhman:2014b}.
$R_i$ is the position of the Earth for WISE observations;
for Spitzer observations, $R_i$ is the position of the spacecraft.
The observed positional difference on the left hand side is in arcsec,
the parameters $\Delta\alpha$ and $\Delta\delta$ are in arcsec,
the proper motion $\mu_\alpha$ and $\mu_\delta$ are in arcsec/yr,
and the parallax $\varpi$ is in arcsec.

In 2010, $x_i$ and $y_i$ are small, and the value for the parameter $f$
is close to one, so the non-standard terms $(f-1)x_i$ and $(f-1)y_i$ on the
right hand side of Eqn \ref{eqn:modeq} do not cause problems.
But they do require inflating the standard deviations of the 2010 data points
by a factor of $f$, since a change of $x$ by $f\sigma$ is necessary to
change the left hand side of Eqn \ref{eqn:modeq} by $\sigma$.
An iterative cycle of fitting for $f$, reweighting, then refitting for $f$ converges
rapidly.
In addition to this error inflation, there is an error with the form 
$(f-1)\sigma_\mathrm{bkgnd}$ which is perfectly correlated across all
the 2010 data.
The background clump is faint  in W2 leading to an imprecise position,
but the W2 only position is not affected by color differences among the clump
members.
The effects due to the 
uncertainty in the clump position, the $(f-1)\sigma_\mathrm{bkgnd}$ term,
are significant and need to be added in quadrature to the errors from the least squares
fit.  
This was done by running the solutions for three background clump
positions: the nominal position, a position with $+1\sigma_\delta$, and
a position with $+1\sigma_\alpha$.   Differences between these solutions
give the sensitivity of the parameters to the background clump position
errors.  The final uncertainty on a parameter, such as $\varpi$, is
given by $\sigma(\varpi)^2 = \sigma(\mathrm{LSQ})^2 + 
(\sigma(\delta_\mathrm{bkgnd}) \partial\varpi/\partial \delta_\mathrm{bkgnd})^2 +
(\sigma(\alpha_\mathrm{bkgnd}) \partial\varpi/\partial \alpha_\mathrm{bkgnd})^2$,
where $\sigma(\mathrm{LSQ})$ is the usual error reported by the least squares fit.
The resulting parameter values and uncertainties are
$\mu_\alpha= -8.051 \pm 0.047$ \asec/yr,
$\mu_\delta = 0.657 \pm 0.050$ \asec/yr,
parallax $= 448 \pm 33$ mas, and
$f = 1.237 \pm 0.071$.  For this fit $\chi^2 = 2.59$ with 4 degrees of
freedom. Figure \ref{fig:astrom} shows the data and the fit with
the proper motion removed.

W0855 passed from East to West of the
background clump during 2010, and the mean
right ascension during 2010 is well determined.
As a result
our $\mu_\alpha$ differs from Luhman's
value by only 9 mas/yr, which supports our reduced
uncertainty on $\mu_\alpha$.  The proper motion in
right ascension is highly correlated with the parallax,
so by reducing the uncertainty in $\mu_\alpha$ we also
get a reduced uncertainty in the parallax.
The difference between this paper and  \citet{luhman:2014b} in $\mu_\delta$ is
$43 \pm 86$ mas/yr when we use the W2 only position for the clump of
background sources.  

The value of $f$ implies that the centroid is
a weighted sum of W0855 and the background source clump
with weights $1/f = 81\pm4\,\%$ for W0855 and
$19\mp4\,\%$ for the background source clump.
The total W2 flux is divided with $79\pm2.5\,\%$
from W0855 and $21\mp2.5\,\%$ from the background
source clump.  Thus the derived value for $f$ is reasonable.

Our conclusion about the astrometry of W0855 is that Luhman's
procedure for correcting the 2010 data for the effect of the background
source clump was quite successful, and our new data and independent
analysis technique give a parallax only 6 mas different than
\citet{luhman:2014b}.   Further astrometry of W0855 could reduce
our reliance on the confused WISE 2010 data, and
this may come from
NEOWISE-R which should get 5 more epochs, from further Spitzer
observations, or from HST detections of the J or H band flux.

\section{Discussion}

The $\geq$Y2 brown dwarf 
W1828+2650 is so faint at J and H that the HST is needed to
measure an accurate color: F140W-W2 $= 8.76 \pm 0.11$ mag for this
very red object.  But
W1828+2650 is not extremely low luminosity given its 
absolute magnitude $M_{W2} = 14.17 \pm 0.25$
\citep{beichman/etal:2013}.
The distant white dwarf companion WD0806-661B has F125W-[4.5] $=
8.81 \pm 0.14$ mag \citep{gelino/etal:2014, luhman/etal:2012} but an
absolute magnitude $M_{4.5} = 15.46 \pm 0.07$ \citep{luhman/etal:2012}.
On the other hand the Y1 brown dwarf 
W0350-5658 is bluer, with F140W-W2 $= 7.57 \pm 0.21$ mag,
but much less luminous with $M_{W2} = 17.05 \pm 0.38$ \citep{marsh/etal:2013}.
Thus W0855, with $M_{W2} = 17.27 \pm 0.17$, is less luminous than the least
luminous previously known brown dwarfs, 
and as red or redder than the reddest previously known brown dwarfs.
Many more objects redder than W0855 or less luminous than W0855
are needed to understand the trend and scatter of the luminosity \vs\
color relation for these very low luminosity objects.

We are fortunate that
W0855 is approximately 2 magnitudes above the AllWISE catalog limit,
so a full search of the entire catalog would cover a volume
$(10^{0.8})^{1.5} = 15.85$ larger than the volume of the sphere
containing W0855.  Thus there is a good possibility that many more
examples of W0855-like objects exist in the AllWISE catalog,
but in the absence of a third epoch to confirm the existence
of a moving W2 only source it is not practical to obtain followup data
from Spitzer or any ground-based telescope.  The ``statistics of one''
limit the precision of any estimated number, but \citet{kerman:2011}
recommends using a $\lambda^{-2/3}$ prior for the Poisson rate $\lambda$,
which gives a posterior probability density $\propto \lambda^{1/3} \exp(-\lambda)$
when $n=1$ objects have been seen.  This posterior has its $16^{th}$ \%-tile
at $0.33$, its median at $n=1.01$, and its $84^{th}$ \%-tile at 2.33.
With this posterior we estimate the number
of objects that a survey of 15.85 times more volume would see, again using a
Poisson distribution.  This calculation gives a $16^{th}$ to $84^{th}$ \%-tile
range of 4 to 35, with a median of 15.

The combination of NEOWISE-R data with the AllWISE database will provide
the additional observation epochs needed to confirm AllWISE motion
detections of low SNR sources, 
making it practical to find more  W0855-like objects if they exist.

\section{Conclusion}

Early data from NEOWISE-R have provided new astrometric
and photometric parameters for W0855.  These are all
consistent with the results of \citet{luhman:2014b}, and confirm
that W0855 is an extremely red object with an extremely faint
absolute magnitude $M_{W2} = 17.27 \pm 0.17$.  W0855 shows
that a population of very cold brown dwarfs exists that are so red that
they are effectively unobservable from the ground at near-IR wavelengths.
Since W0855 is two magnitudes brighter than the detection limit of WISE
in the W2 band, an astrometric analysis of all the NEOWISE-R
data down to the catalog limit could lead to the discovery of  4-35
similar objects.

\acknowledgements
This publication makes use of data products from the Wide-field
Infrared Survey Explorer, which is a joint project of the University
of California, Los Angeles, and the Jet Propulsion Laboratory/California
Institute of Technology, funded by the National Aeronautics and
Space Administration.  This publication makes use of data products 
from NEOWISE, which is a project of the 
Jet Propulsion Laboratory/California Institute of Technology, 
funded by the National Aeronautics and Space Administration.

{\it Facilities:} \facility{WISE},
\facility{Keck/NIRC2-LGSAO}.

% \bibliography{apj-jour,W0855}

\end{document}